\documentclass[9pt,conference]{IEEEtran}
\usepackage{amsmath,graphicx,algorithm,algorithmic,adjustbox,setspace}

\usepackage{amsfonts}
\usepackage{textcomp}

\usepackage{xcolor}
\usepackage{enumitem}
\usepackage{hyperref}
\usepackage{multirow}
\usepackage{etoolbox,siunitx}
\robustify\bfseries
\usepackage{booktabs}
\usepackage{balance}
\usepackage{amssymb}
\usepackage{arydshln}
\usepackage{bm}
\usepackage{dsfont}
\usepackage{cite}

\usepackage[hang,flushmargin]{footmisc}

\usepackage[inkscapelatex=false]{svg}

\def\R{{\mathbb R}}

\title{Task-Aware Unified Source Separation
\thanks{This work was performed while K.~Saijo was an intern at MERL.}}

\author{\IEEEauthorblockN{\textit{Kohei Saijo$^{1,2}$, Janek Ebbers$^{1}$, François G.\ Germain$^{1}$, Gordon Wichern$^{1}$, Jonathan Le Roux$^{1}$}\thanks{This work was performed while K.~Saijo was an intern at MERL.}\vspace{.7\baselineskip}}        
\IEEEauthorblockA{{$^{1}$Mitsubishi Electric Research Laboratories (MERL), Cambridge, MA, USA} \;
{$^{2}$Waseda University, Tokyo, Japan}}
}

\begin{document}
\maketitle
\begin{abstract}
Several attempts have been made to handle multiple source separation tasks such as speech enhancement, speech separation, sound event separation, music source separation (MSS), or cinematic audio source separation (CASS) with a single model.
These models are trained on large-scale data including speech, instruments, or sound events and can often successfully separate a wide range of sources.
However, it is still challenging for such models to cover all separation tasks because some of them are contradictory (e.g., musical instruments are separated in MSS while they have to be grouped in CASS).
To overcome this issue and support all the major separation tasks, we propose a task-aware unified source separation (TUSS) model.
The model uses a variable number of learnable prompts to specify which source to separate, and changes its behavior depending on the given prompts, enabling it to handle all the major separation tasks including contradictory ones. 
Experimental results demonstrate that the proposed TUSS model successfully handles the five major separation tasks mentioned earlier.
We also provide some audio examples, including both synthetic mixtures and real recordings, to demonstrate how flexibly the TUSS model changes its behavior at inference depending on the prompts.

\end{abstract}

\begin{IEEEkeywords}
Unified source separation, %
prompts, task-aware
\end{IEEEkeywords}
\section{Introduction}
\label{sec:intro}

With the advent of neural network-based approaches, high-fidelity audio source separation systems have been developed for multiple applications.
Source separation has historically been formulated as one of several sub-tasks, such as speech enhancement (SE)~\cite{WDL2018, reddy2020interspeech,zhang2023toward}, speech separation (SS)~\cite{dc, pit,convtasnet,wavesplit,dprnn,sepformer,tfgridnet}, music source separation (MSS)~\cite{stoter19,sawata2021all,bsrnn}, and universal sound separation (USS)~\cite{universal_sound_separation,tzinis2020improving}.
Recently, the task of separating a mixture into the broader categories of speech, music, and sound effects (SFX) was introduced as cinematic audio source separation (CASS), also known as the cocktail fork problem~\cite{zhang2021multi,petermann2022cocktail,Uhlich2024CDX}.
In some cases, all the sources in a mixture need to be separated, while in others the desired stems may themselves be mixtures of multiple sources, such as in CASS, the noise stem in SE, or the {\it others} stem in MSS. %
In most cases, separation models are trained on specific datasets and address only a specific type of task.

In contrast, the recently proposed general audio source separation (GASS)~\cite{pons2024gass} aims to
develop a single model that can separate arbitrary sources\footnote{While USS~\cite{universal_sound_separation} originally aims to separate arbitrary sources, it has so far been mostly limited to the separation of predominantly sound event sources. Following~\cite{pons2024gass}, we use the term ``GASS" for the separation of mixtures that may contain speech, music, and/or sound events.}.
While single separation models that can separate speech, musical instruments, and environmental sounds well could be obtained by training on large-scale data, the models had a fixed number of outputs and needed to be fine-tuned on each downstream task to reach satisfactory performance.
We argue that this is because the source separation problem is inherently ill-posed and its goal is task-specific. In particular, it is challenging for a single task-agnostic model such as in GASS to handle tasks with contradictory goals (e.g., CASS where music sources need to be grouped and MSS where they need to be separated), as it cannot know what source to separate.

To handle such contradictory tasks, a potential approach would be to develop a conditional separation model which would change its behavior depending on the given condition. Conditional models have been mainly developed so far for target sound extraction (TSE), 
specifying a target source using a cue such as a speaker utterance or sound recording~\cite{vzmolikova2019speakerbeam,wang2022few,chen2022zero}, 
or specifying a target sound event class (or group thereof) using class IDs~\cite{wang19h_interspeech, seetharaman2019class, ochiai20_interspeech,tzinis2022heterogeneous,delcroix2022soundbeam}. %
In particular, text-queried TSE models~\cite{kilgour22_interspeech, liu22w_interspeech}, where the source or group of sources to extract is specified by a natural language prompt, might be considered as a potential way to handle all the tasks mentioned above.
However, unlike normal unconditional separation models, TSE models extract only one source or group of sources, and they do not explicitly model the relationship between the target source and the other sources.

To go beyond these limitations and truly address all the major source separation tasks mentioned earlier and potentially others, we propose a task-unified source separation model.
The model has learnable prompts, corresponding to speech, SFX, SFX-mix, and so on, and separates sources specified by the prompts.
Unlike traditional source separation and TSE models, the proposed model accepts a variable number of prompts and outputs simultaneously the corresponding number of separated sources. This allows the model to use the information from other prompts and to handle the separation of multiple sources from the same class beyond the classical speech separation case.
The model features prompts to obtain an individual source (e.g., SFX) as well as a mixture of sources (e.g., SFX-mix), which allows it to handle all the tasks including CASS.

In our experiments, we demonstrate that the proposed model successfully handles multiple tasks with a single model, allowing a user to flexibly control the desired outputs for a given mixture at inference time.
Informal testing also shows that the model is able to handle combinations of prompts unseen during training.

\begin{figure}[t]
\centering
\centerline{\includegraphics[width=0.8\linewidth]{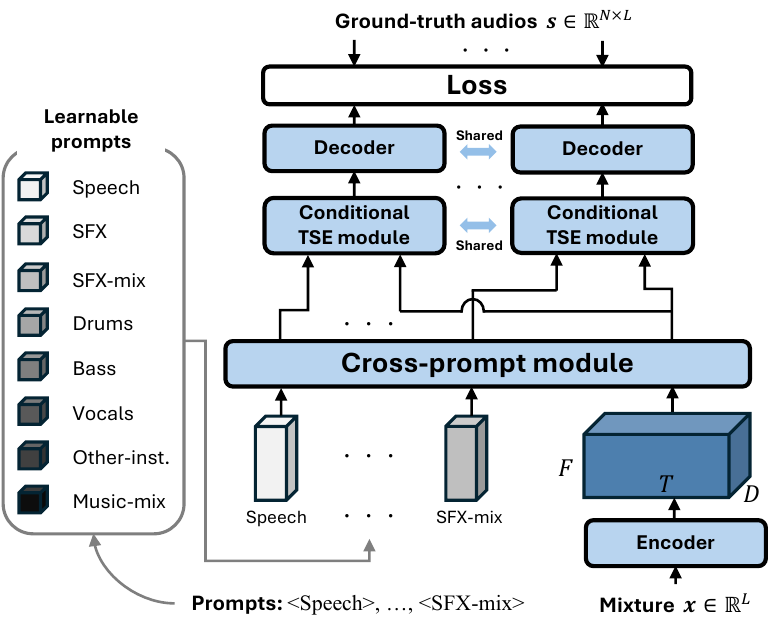}}\vspace{-2mm}
\caption{
   Overview of the proposed task-aware unified source separation model.
   Receiving the input mixture's encoded feature and learnable prompts that specify which sources to separate, the cross-prompt module first jointly models both as a sequence to condition one on the other.
   Then, the source specified by each prompt is extracted by the conditional TSE module.
   $N$ sources are separated given $N$ prompts, where $N$ can be a variable number.
}
\label{fig:overview}
\vspace{-3mm}
\end{figure}

\section{Task-aware unified source separation}
\label{sec:prop}

\subsection{Problem setup}
\label{ssec:problem_setup}

The goal of this work is to build a \textit{unified} source separation model that addresses all the major source separation tasks such as SE, SS, USS, MSS, and CASS.
Since some tasks have contradictory goals, we believe that we need a conditional separation model that can change its behavior, including the number of output sources, depending on the condition.
To this end, we propose a task-aware unified source separation (TUSS) model whose behavior is controlled by several learnable prompts to specify what source to separate, as shown in Fig.~\ref{fig:overview}.
Specifically, we split sources into the following 8 categories and prepare the corresponding prompts: \texttt{<Speech>}, \texttt{<SFX>}, \texttt{<SFX-mix>}, \texttt{<Drums>}, \texttt{<Bass>}, \texttt{<Vocals>}, \texttt{<Other inst.>}, and \texttt{<Music-mix>}.
The \texttt{<*-mix>} prompts are for grouping all the sources from that category, while the others are for extracting individual sources.
As shown in Table~\ref{table:task_and_prompts}, the five typical tasks mentioned earlier can be covered by changing the combination of the prompts.
The model also accepts other arbitrary combinations of prompts, except for the combinations including both \texttt{<SFX-mix>} and \texttt{<SFX>}, and \texttt{<MUSIC-mix>} and individual instruments. More prompts could of course be added in the future to handle a greater variety of tasks. In particular, we did not include a \texttt{<Speech-mix>} prompt for extracting speech mixtures as this is not a conventional task, but we readily could.

To address all five tasks in Table~\ref{table:task_and_prompts}, the model has to satisfy the following requirements: i) a variable number of prompts are acceptable since each task has a different number of outputs, and ii) multiple identical prompts are acceptable (e.g., $N$-speaker SS is specified via $N$ \texttt{<Speech>} prompts, all identical, and the model has to output $N$ different speech signals).
The proposed TUSS model satisfies both requirements by using a Transformer-based architecture. %

\begin{table}[t]
\centering
\sisetup{
detect-weight, %
mode=text, %
tight-spacing=true,
round-mode=places,
round-precision=1,
table-format=2.1,
table-number-alignment=center
}
\caption{
    Tasks and corresponding prompts. $\times N$ means that the same prompt is repeated $N$ times.
}
\vspace{-0.1in}
\label{table:task_and_prompts}
\resizebox{0.9\linewidth}{!}{
\begin{tabular}{ll}

\toprule
Task  &Prompts \\

\midrule

SE & \texttt{<Speech>}, \texttt{<SFX-mix>} \\

(noisy-) SS & \texttt{<Speech>} $\times N$, (\texttt{<SFX-mix>}) \\

USS & \texttt{<SFX>} $\times N$ \\

MSS & \texttt{<Drums>}, \texttt{<Bass>}, \texttt{<Vocals>}, \texttt{<Other inst.>} \\

CASS & \texttt{<Speech>}, \texttt{<SFX-mix>}, \texttt{<MUSIC-mix>} \\

\bottomrule

\end{tabular}
}
\vspace{-3mm}
\end{table}

\subsection{Task-aware unified source separation model}
\label{ssec:tuss_model}

An overview of the TUSS model is shown in Fig.~\ref{fig:overview}.
The model comprises an encoder, learnable prompts, a cross-prompt  
module, a conditional TSE 
module, and a decoder.
Learnable prompts are initialized randomly and jointly trained with the separation model.

The \textbf{encoder} first applies the short-time Fourier transform (STFT) to the time-domain waveform $\bm{x} \in \R^{L}$ ($L$ is the number of samples), resulting in a time-frequency (TF-)domain representation $\bm{X} \in \R^{2 \times T \times F}$, where $T$ is the number of frames, $F$ that of frequency bins, and 2 corresponds to real and imaginary parts.
$\bm{X}$ is further encoded using a learnable linear layer (detailed in Section~\ref{ssec:model}), resulting in a 3-d tensor $\bm{Z} \in \R^{D \times T \times F}$.

The \textbf{cross-prompt module} is the core processing part to achieve the two requirements mentioned in Section~\ref{ssec:problem_setup}.
$N$ learnable prompts $\bm{P}_n$ (each with shape $D \times 1 \times 1$) are first stacked $F$ times along the frequency dimension and then concatenated at the front of the encoded feature $\bm{Z}$ along the temporal dimension, resulting in a tensor $\bm{Z}' = [\bm{P}, \bm{Z}] \in \R^{D \times (N+T) \times F}$.%
$\bm{Z}'$ is then input to Transformer-based blocks to model the dependency of the temporal sequence.
This process not only enables the mixture to be modeled conditioned by the prompts but also allows each prompt to be processed conditioned on the mixture and the other prompts, which helps the conditional separation in the conditional TSE module.
Thanks to positional encoding and self-attention, even identical prompts at different positions result in different values.
In addition, the Transformer-based architecture by design accepts sequences with arbitrary length, which enables the model to receive any number of prompts. %

The \textbf{conditional TSE module} extracts the source specified by each prompt in parallel.
The output $\tilde{\bm{Z}}'$ of the cross-prompt module is first split into the features $\tilde{\bm{P}}_n$ corresponding to each prompt and the feature $\tilde{\bm{Z}}$ corresponding to the mixture.
Then each prompt $\tilde{\bm{P}}_n$ is multiplied (with broadcasting) by $\tilde{\bm{Z}}$, resulting in a feature conditioned by a prompt $\tilde{\bm{Z}}_{n} = \tilde{\bm{Z}} \odot \tilde{\bm{P}}_n$.
Each $\tilde{\bm{Z}}_{n}$ is further processed by several learnable layers, which are shared for all $n$.

The \textbf{decoder} receives each output $\hat{\bm{Z}}_{n}$ of the conditional TSE module as input, and converts it back to the time-domain waveform using an MLP block and inverse STFT, resulting in separated signals $\hat{\bm{s}} \in \R^{N \times L}$. The decoder is also shared for all $n$.

When computing the loss, although the order of the separated signals is the same as that of the prompts, we do not know the order of sources when multiple prompts from the same category are used.
We thus compute the permutation-invariant (PIT) loss~\cite{dc, pit} for each category independently and average the loss of each category.

\subsection{Prompt dropout}
\label{ssec:prompt_dropout}
In Section~\ref{ssec:tuss_model}, we assume that $N$ prompts are input to separate all $N$ sources in a mixture.
However, in practice, a user may sometimes wish to separate only a subset of sources.
To handle this case, we introduced prompt dropout, where $M$ prompts ($M < N$) are removed and the model tries to separate only $N - M$ sources during training.
Specifically, in 25\% of the training steps, we uniformly sample $M$ from $[1, N)$ and remove $M$ prompts randomly.
Here, when the prompts include multiple prompts from the same category, we do not remove them because the model would have no objective way to know which of the sources from that category to separate. %

\section{Related work}
\label{sec:related_work}

Several attempts have been made to build multi-task source separation models.
In~\cite{saijo2023single}, a single model that supports five SE/SS tasks has been proposed.
Closest to our work, GASS aims to separate arbitrary sources by training a model on large-scale data~\cite{pons2024gass}.
Again, it is challenging for GASS by design to support contradictory tasks such as MSS and CASS.

One way to handle contradictory tasks is via hierarchical separation, where the model has multiple prediction heads, for example, to estimate category-wise mixtures and individual sources~\cite{manilow2020hierarchical, petermann2023hyperbolic}.
While they have a fixed number of outputs for individual sources (just one source~\cite{manilow2020hierarchical} or one for each category~\cite{petermann2023hyperbolic}), TUSS can change the number of outputs for each category.
While hierarchy is not explicitly enforced in TUSS, introducing it for example via hyperbolic prompt embeddings is an interesting avenue for future research.

In~\cite{zhang2023toward}, a learnable prompt is used to specify whether the model should perform dereverberation in an SE model.
Although such a model and our TUSS model are similar in that the model's behavior can be changed with prompts, our model supports multiple prompts combined in arbitrary order and number, which enables it to support multiple separation tasks.

\section{Experiments}
\label{sec:experiments}

\subsection{Datasets}
\label{ssec:datasets}

During training, we create mixtures on the fly using the datasets in Table~\ref{table:datasets}.
LibriVox data is from the URGENT challenge, where DNSMOS-based filtering is done to remove some noisy speech~\cite{zhang2024urgent}.
For FSD50K, we first filter out human speech and musical instruments.
We then split them into two groups, ``single" and ``multi", depending on the number of leaf sound-class labels and the audio length, following a similar procedure to~\cite{fuss, pons2024gass}.
``Single" includes audio with a single sound-class label and shorter than 8 s, while ``multi" includes those with multiple labels or longer than 8 s.

When creating a mixture, the number of prompts $N$ is first randomly sampled from $2$ to $4$, and then
$N$ prompts are selected\footnote{
We draw a more realistic combination of prompts more frequently (e.g., Bass and Drums co-occur more often than Bass and Speech). We will publish the code and detailed configuration once the paper is accepted.}, where \texttt{<SFX-mix>} and \texttt{<SFX>} or \texttt{<MUSIC-mix>} and individual instruments cannot coexist in a mixture.
Here, \texttt{<Speech>} and \texttt{<SFX>} can be selected multiple times while the others can be chosen only once.
Next, an audio file from the corresponding category is randomly sampled from the datasets in Table~\ref{table:datasets} for each prompt.
For SFX-mix and Music-mix, we sometimes mix multiple sources from SFX or Music Inst.\ on the fly, instead of using FSD50K or FMA.
Since sources from different datasets can have different sampling rates, we re-sample the sources to the lowest sampling rate among selected sources, then up-sample them to 48 kHz.
Finally, sources are RMS-normalized, scaled by gains uniformly sampled from the ranges shown in Table~\ref{table:datasets}, %
and mixed to create a mixture.

\begin{table}[t]
\centering
\sisetup{
detect-weight, %
mode=text, %
tight-spacing=true,
round-mode=places,
round-precision=1,
table-format=2.1,
table-number-alignment=center
}
\caption{
    Datasets used during training in each category.
    We split FSD50K into ``single" and ``multi", depending on the number of sound event labels.
    Datasets with $\dagger$ are mixed on the fly to create the SFX-mix and Music-mix categories.
}
\vspace{-0.1in}
\label{table:datasets}
\resizebox{\linewidth}{!}{
\begin{tabular}{lcl}

\toprule
Category &Gain~[dB]  &Datasets \\
\midrule

Speech      &[-10, 0]  & VCTK~\cite{veaux2013voice}, WSJ0~\cite{wsj0}, LibriVox from URGENT challenge~\cite{zhang2024urgent} \\
SFX         &[-10, 0]  & FSD50K-single~\cite{fonseca2021fsd50k} \\
SFX-mix     &[-20, 0]  & WHAM!~\cite{wichern19_interspeech}, DEMAND~\cite{thiemann2013diverse}, FSD50K-multi, FSD50K-single$^\dagger$ \\
Music Inst. &[-10, 0]  & MUSDB-HQ~\cite{MUSDB18HQ}, MOISESDB~\cite{pereira2023moisesdb} \\ 
Music-mix   &[-20, 0]  & FMA~\cite{defferrard2016fma}, MUSDB-HQ$^\dagger$, MOISESDB$^\dagger$ \\
\bottomrule

\end{tabular}
}
\vspace{-2mm}
\end{table}

We employ the evaluation partition of five datasets to evaluate our model on multiple separation tasks.
\textbf{VCTK-DEMAND} is used for the SE task. It includes noisy speech mixtures derived from VCTK speech and DEMAND noise sampled at 48 kHz.
\textbf{WHAM!}
(max version) is used for the noisy SS task. Speech and noise are from the WSJ and WHAM! corpora, respectively, sampled at 16 kHz.
\textbf{FUSS} is used for the USS task. Two to four sources from the FSD50K corpus sampled at 16kHz are mixed. Note that we removed single-source mixtures from the original FUSS dataset.
\textbf{MUSDB-HQ} is used for the MSS task, where the goal is to separate mixtures into vocals, bass, drums, and other instruments. The sampling rate is 44.1 kHz.
\textbf{DnR} is used for the CASS task. Speech, Music-mix, and SFX-mix sources are obtained from LibriSpeech, free music archive (FMA), and FSD50K, respectively, sampled at 44.1 kHz.

\subsection{Model Architecture}
\label{ssec:model}

Both the cross-prompt module and the conditional TSE module consist of several TF-Locoformer blocks~\cite{tflocoformer}.
Each TF-Locoformer block has frequency modeling and temporal modeling sub-blocks, each based on multi-head self-attention and convolutional feed-forward networks (see \cite{tflocoformer} for more details).
While the original TF-Locoformer features an STFT+conv2d encoder and a deconv2d+iSTFT decoder, we replace here the conv2d and deconv2d with a band-split encoder and a band-wise decoding module~\cite{bsrnn} to efficiently handle data with high sampling rates.
The band-split encoder splits the input spectrogram $\bm{X}$ with $F$ frequency bins into $K$ subband spectrograms $\bm{X}_{k}$ $(k={1,\dots ,K})$ with pre-defined band-widths $b_{k}$ satisfying $\sum_{k} b_{k} = F$.
The real and imaginary parts of each subband spectrogram are concatenated as described in Section~\ref{ssec:tuss_model} and processed with a normalization layer and a linear layer, resulting in a feature $\bm{Z}_{k}$.
The $K$ features are then concatenated and result in a feature with shape $T \times K \times D$, which is processed by TF-Locoformer blocks.
The band-wise decoding module again splits the feature into $K$ sub-features and decodes them to obtain band-wise masks (see ~\cite{bsrnn} for more details).
We follow the same band-split configuration as~\cite{bsroformer} with $K=62$ bands.

We train Medium and Large models.
For the Medium model, we set $B=4$, $D=64$, $C=384$, $K=4$, $S=1$, $H=4$, $G=8$, and attention hidden size $E$ of 256 in the cross-prompt module (notations follow the TF-Locoformer paper~\cite{tflocoformer}).
In the conditional TSE module, we use the same settings, except for $B=2$, $C=384$, and $E=96$.
For the Large model, we set $B=6$, $D=128$, $C=384$, $K=4$, $S=1$, $E=256$ $H=8$, and $G=8$ in the cross-prompt module, and $B=3$, $C=256$, and $E=192$ with other settings unchanged in the conditional TSE module.
The Medium and Large models have 11.1M and 38.2M parameters, respectively.
Note that a linear layer is used instead of a convolution layer for the temporal modeling in the cross-prompt module, while a convolution layer is used in the conditional TSE module.

\subsection{Compared methods}
\label{ssec:compared_methods}

\begin{table*}[t]
\centering
\sisetup{
detect-weight, %
mode=text, %
tight-spacing=true,
round-mode=places,
round-precision=1,
table-format=2.1,
table-number-alignment=center
}
\caption{
    Evaluation results of medium (\texttt{M*}) and large (\texttt{L*}) models. SNR [dB] is shown for MUSDB-HQ and SI-SNR [dB] for other test sets.
    The model  of \texttt{L6} and \texttt{L6s} is fine-tuned using prompt dropout (Section~\ref{ssec:prompt_dropout}).
}
\vspace{-0.1in}
\label{table:results}
\resizebox{\linewidth}{!}{
\begin{tabular}{llc*{12}{S}}

\toprule

& &\multirow{2}{*}[-1.55ex]{\shortstack{Eval.\\Prompts}} &\multicolumn{2}{c}{VCTK-DEMAND (SE)} &\multicolumn{2}{c}{WHAM! (SS)} &{FUSS (USS)} &\multicolumn{4}{c}{MUSDB-HQ (MSS)} &\multicolumn{3}{c}{DnR (CASS)} \\
\cmidrule(lr){4-5}\cmidrule(lr){6-7}\cmidrule(lr){8-8}\cmidrule(lr){9-12}\cmidrule(lr){13-15}

& & &{Speech} &{SFX-mix} &{Speech} &{SFX-mix} &{SFX} &{Vocals} &{Bass} &{Drums} &{Other} &{Speech} &{Music-mix} &{SFX-mix} \\

\midrule

\texttt{M0}&Conventional data specialist &- &17.4 &8.5 &\bfseries 9.3 &\bfseries 12.3 &8.3 &9.1 &6.5 &9.3 &5.6 &\bfseries 15.1 &\bfseries 7.0 &\bfseries 8.3  \\
\texttt{M1}&Prompting data specialist  &all &17.6 &9.0 &\bfseries 9.3 &12.1 &\bfseries 10.2 &9.4 &7.1 &9.0 &5.9 &14.8 &6.9 &8.2  \\

\texttt{M2}&Conventional task specialist &- &\bfseries 20.4 &\bfseries11.6 &8.2 &11.9 &8.3 &9.4 &7.4 &9.9 &\bfseries 6.0 &14.6 &6.1 &7.4  \\
\texttt{M3}&Prompting task specialist &all &20.2 &\bfseries 11.6 &8.5 &12.0 &\bfseries 10.2 &\bfseries 9.6 &\bfseries 7.6 &\bfseries 10.2 &5.9 &14.9 &6.6 &7.7  \\

\texttt{M4}&Conventional unified &- &19.2 &8.7 &3.4 & 10.2 &8.1 &7.2 &2.5 &6.6 &3.1 &12.0 &3.8 &2.1  \\
\texttt{M5}&Prompting unified &all &19.4 &10.2 &7.0 &11.2 &9.6 &8.4 &5.7 &8.3 &4.7 &14.5 &5.7 &7.1  \\

\midrule

\texttt{L3}&Prompting task specialist &all & \bfseries 20.4 &\bfseries  11.5 &\bfseries 10.1 &\bfseries 12.6 &10.0 &\bfseries 10.4 &\bfseries 8.4 &\bfseries 11.1 &\bfseries 6.5 &\bfseries 15.1 &\bfseries 7.0 &7.9  \\

\texttt{L4}&Conventional unified &- &19.2 &9.1 &6.5 &10.8 &9.9 &7.7 &4.5 &7.5 &4.2 &12.5 &5.6 &4.9  \\

\texttt{L5}&Prompting unified &all &19.8 &10.4 &8.7 &11.9 &\bfseries 12.2 &8.6 &6.3 &9.1 &5.4 &\bfseries 15.1 &\bfseries 7.0 &\bfseries 8.2  \\

\texttt{L5s}\!&Prompting unified &single &16.2 &10.2 &8.7 &6.5 &\bfseries 12.2 &-6.5 &-3.0 &-2.9 &-5.0 &13.0 &-1.5 &0.4  \\

\texttt{L6}&Prompting unified (fine-tuned)\! & all &19.6 &9.8 &8.8 &11.8 &9.0 &7.3 &4.3 &8.3 &3.4 &14.9 &6.5 &7.8  \\

\texttt{L6s}\!&Prompting unified (fine-tuned)\! & single &18.7 &9.8 &8.8 &10.4 &9.0 &6.5 &3.9 &7.9 &2.4 &14.5 &3.8 &7.2  \\

\bottomrule

\end{tabular}
 }
\vspace{-4mm}
\end{table*}

To assess how well the unified model handles all the tasks, we train specialist models for each task as the baseline.
We train two types of specialists, \textit{data specialist} and \textit{task specialist}.
The data specialist is trained using the same data source as the test set (e.g., WSJ0 speech and WHAM! noise are used for the WHAM! data specialist model), while the task specialist uses all the data for that task (e.g., all the speech and SFX-mix data are used on top of WHAM! and VCTK-DEMAND for the SE task specialist model). We refer to all our prompt-based models as \emph{prompting} models in the results.

We also train \emph{conventional} separation models that do not have any prompts and output a fixed number of sources, as in~\cite{fuss, pons2024gass}, using an architecture as close as possible to ours for fair comparison (TF-Locoformer is the current state of the art on all the tasks it has been tested on).
Specifically, the encoded feature goes through some TF-Locoformer processing blocks and the decoder estimates multiple outputs from the processed feature.
Since different mixtures have different numbers of sources, the model has four outputs and is trained to output zeros when the number of sources in the input mixture is fewer than four.
We also train data and task specialist versions of the conventional model, where the number of outputs is the same as that of sources in the mixture.
Note that the number of sources randomly changes from two to four in each training step when training the FUSS specialist models, following its task definition~\cite{fuss}.

\subsection{Training and evaluation details}
\label{ssec:train_eval_details}

We train the models for 150 epochs, where 1 epoch is 2.5k training steps.
We use the AdamW optimizer~\cite{adamw} with a weight decay factor of 1e-2.
The learning rate is linearly increased from 0 to 1e-3 in the Medium model and 5e-4 in the Large model for the first 10k steps, kept constant for 75 epochs, and then decayed by 0.5 if the validation loss does not improve for 5 epochs.
Gradient clipping is applied with a maximum gradient $L_2$-norm of 5.
The batch size is 8 and the input mixture is 6~\si{\second} long.
When training with prompt dropout, we initialize the model using the parameters at the 124-th epoch and fine-tune it for 26 epochs to save training time.
Configuration for the optimization and the learning rate schedule is the same as above, except that the peak learning rate is 1.25e-4.
The negative signal-to-distortion ratio (SNR) is used as the loss function.
For the conventional model, we use the SNR loss that accepts zero signals as the ground truth~\cite{fuss} so that the model can handle mixtures with fewer than four sources.

\subsection{Main results}
\label{ssec:results}

Table~\ref{table:results} shows the evaluation results of the conventional model and the proposed TUSS model trained on each dataset (data specialist), all data for each task (task specialist), and all the combined data (unified).
\texttt{M*} and \texttt{L*} respectively indicate Medium and Large models.
Note that the performance is the same for the data specialist and task specialist on FUSS because we only use FSD50K as SFX.

First, comparing the conventional and prompting specialist models (\texttt{M0} vs.\ \texttt{M1} and \texttt{M2} vs.\ \texttt{M3}), they achieve comparable performance for all the tasks, which validates that the design of the TUSS model does not harm the performance.
Note that the TUSS model is better on FUSS but it is not a fully fair comparison since TUSS assumes the number of sources is known.
Interestingly, although task specialists leverage more data, their performance is inferior to data specialists on WHAM! and DnR! test sets (\texttt{M0} vs.\ \texttt{M2} and \texttt{M1} vs.\ \texttt{M3}).
We believe this is likely because of the domain mismatch between training and inference.
For instance, SFX-mix data in DnR are mainly composed of multiple short environmental sounds but task specialists also use other SFX-mix data such as WHAM! or DEMAND that are more akin to background noise.
In the future, we will address this issue by e.g., separating SFX-mix and background noise categories.

Models \texttt{M4} and \texttt{M5} are trained on all the datasets.
The results show that the proposed TUSS model \texttt{M5} better addresses all the tasks, validating our hypothesis that conditional models like TUSS are more appropriate to handle multiple tasks, including contradictory ones.
Compared with the specialist models, the unified TUSS model does not improve performance (e.g., \texttt{M3} vs.\ \texttt{M5}).
However, since the unified model can utilize larger-scale data, it may benefit from a larger model~\cite{zhang2024beyond}.
Indeed, comparing \texttt{L3} vs.\ \texttt{L5}, the TUSS model outperforms the specialist model on some datasets (FUSS and DnR).
Although the TUSS model still falls behind the specialist model on some datasets, the results imply that TUSS may eventually outperform specialists by carefully scaling the data and model.

While \texttt{L5} assumes that all the prompts are input to separate all the sources in a mixture, we fine-tuned \texttt{L5} with prompt dropout so that the model can separate a subset of sources (cf.\ Section~\ref{ssec:prompt_dropout}).
The results of the fine-tuned model are \texttt{L6} and \texttt{L6s}, where \texttt{L6} is evaluated using all the prompts while \texttt{L6s} only receives prompts for a single category in each forward pass (e.g., inference on WHAM! is done in two steps, with [\texttt{<Speech>}, \texttt{<Speech>}] and with [\texttt{<SFX-mix>}]).
We applied the same evaluation scheme to the non-fine-tuned \texttt{L5} model and listed the results as \texttt{L5s}.
First, we confirm that prompt-dropout fine-tuning only has a limited impact on the model's performance when using all prompts (\texttt{L5} vs.\ \texttt{L6}).
While the model without prompt-dropout fine-tuning shows a significant performance drop (\texttt{L5} vs.\ \texttt{L5s}), the fine-tuned model maintains relatively good performance with a subset of prompts (\texttt{L5} vs.\ \texttt{L6s}), validating the effectiveness of prompt dropout.

\subsection{Informal test to assess the flexibility at inference}

While the TUSS model falls behind the specialist models on some tasks, TUSS can separate new combinations of sources unseen in the five tasks in Table~\ref{table:results} by changing the prompts, which cannot be achieved by the specialist or conventional models.
To assess such flexibility of the unified model, we conducted an informal test and provide examples on our demo page\footnote{\url{https://www.jonathanleroux.org/research/ICASSP2025-tuss/}}.

For DnR mixtures, for instance, we can consider multiple combinations of prompts depending on whether to separate Music-mix or SFX-mix.
While we normally use [\texttt{<Speech>}, \texttt{<Music-mix>}, \texttt{<SFX-mix>}], the model can separate individual SFX sounds just by changing the prompts to [\texttt{<Speech>}, \texttt{<Music-mix>}, \texttt{<SFX>}, \texttt{<SFX>}].
The model also separates musical instruments well by replacing [\texttt{<Music-mix>}] with, e.g., [\texttt{<Drums>}, \texttt{<Other inst.>}].
These results imply that we can control the model's behavior very easily.

We also conducted the test on FMA, which is always used as MUSIC-mix data during training.
We found that it is challenging for the conventional model \texttt{M4} to separate FMA data, possibly because the model overfits to FMA data as a source that is not to be separated.
In contrast, the TUSS model \texttt{M5} successfully separates all the sources, even though FMA data is also never separated during training, which suggests the advantage of the conditional model over the conventional unconditional model.
We observed a similar trend when testing models on WHAM! noise. On a WHAM! noise containing some faint speech, for example, the conventional model failed to separate speech from noise as 
the WHAM! noise is always assigned to SFX-mix during training, but the TUSS model could separate the two sources well with [\texttt{<Speech>}, \texttt{<SFX>}] prompts.

\section{Conclusion and future work}
\label{sec:conclusion}

This work introduced the Task-aware Unified Source Separation model to address all major separation tasks.
By informing the model of what source to separate using learnable prompts, the model successfully handles multiple tasks.
We also provided some examples that demonstrates the flexibility of the proposed model.
In the future, we plan to support speaker ID and text as prompts via speaker and text embeddings to make the model more versatile.

\clearpage
\balance
\bibliographystyle{IEEEtran}
\bibliography{refs}

\begin{thebibliography}{10}
\providecommand{\url}[1]{#1}
\csname url@samestyle\endcsname
\providecommand{\newblock}{\relax}
\providecommand{\bibinfo}[2]{#2}
\providecommand{\BIBentrySTDinterwordspacing}{\spaceskip=0pt\relax}
\providecommand{\BIBentryALTinterwordstretchfactor}{4}
\providecommand{\BIBentryALTinterwordspacing}{\spaceskip=\fontdimen2\font plus
\BIBentryALTinterwordstretchfactor\fontdimen3\font minus
  \fontdimen4\font\relax}
\providecommand{\BIBforeignlanguage}[2]{{%
\expandafter\ifx\csname l@#1\endcsname\relax
\typeout{** WARNING: IEEEtran.bst: No hyphenation pattern has been}%
\typeout{** loaded for the language `#1'. Using the pattern for}%
\typeout{** the default language instead.}%
\else
\language=\csname l@#1\endcsname
\fi
#2}}
\providecommand{\BIBdecl}{\relax}
\BIBdecl

\bibitem{WDL2018}
D.~Wang and J.~Chen, ``Supervised speech separation based on deep learning: An
  overview,'' \emph{IEEE/ACM Trans. Audio, Speech, Lang. Process.}, vol.~26,
  no.~10, pp. 1702--1726, 2018.

\bibitem{reddy2020interspeech}
C.~K. Reddy, V.~Gopal, R.~Cutler, E.~Beyrami, R.~Cheng, H.~Dubey,
  S.~Matusevych, R.~Aichner, A.~Aazami, S.~Braun \emph{et~al.}, ``The
  {I}nterspeech 2020 {D}eep {N}oise {S}uppression challenge: Datasets,
  subjective testing framework, and challenge results,'' in \emph{Proc.
  Interspeech}, Oct. 2020.

\bibitem{zhang2023toward}
W.~Zhang, K.~Saijo, Z.-Q. Wang, S.~Watanabe, and Y.~Qian, ``Toward universal
  speech enhancement for diverse input conditions,'' in \emph{Proc. ASRU},
  2023, pp. 1--6.

\bibitem{dc}
J.~R. Hershey, Z.~Chen, J.~Le~Roux, and S.~Watanabe, ``Deep clustering:
  Discriminative embeddings for segmentation and separation,'' in \emph{Proc.
  ICASSP}, 2016, pp. 31--35.

\bibitem{pit}
D.~Yu, M.~Kolbæk, Z.~H. Tan, and J.~Jensen, ``Permutation invariant training
  of deep models for speaker-independent multi-talker speech separation,'' in
  \emph{Proc. ICASSP}, 2017, pp. 241--245.

\bibitem{convtasnet}
Y.~Luo and N.~Mesgarani, ``{C}onv-{T}as{N}et: Surpassing ideal time-frequency
  magnitude masking for speech separation,'' \emph{IEEE/ACM Trans. Audio,
  Speech, Lang. Process.}, vol.~27, no.~8, pp. 1256--1266, 2019.

\bibitem{wavesplit}
N.~Zeghidour and D.~Grangier, ``Wavesplit: End-to-end speech separation by
  speaker clustering,'' \emph{IEEE/ACM Trans. Audio, Speech, Lang. Process.},
  vol.~29, pp. 2840--2849, 2021.

\bibitem{dprnn}
Y.~Luo, Z.~Chen, and T.~Yoshioka, ``{Dual-Path RNN}: Efficient long sequence
  modeling for time-domain single-channel speech separation,'' in \emph{Proc.
  ICASSP}, 2020.

\bibitem{sepformer}
C.~Subakan, M.~Ravanelli, S.~Cornell, M.~Bronzi, and J.~Zhong, ``Attention is
  all you need in speech separation,'' in \emph{Proc. ICASSP}, 2021.

\bibitem{tfgridnet}
Z.-Q. Wang, S.~Cornell, S.~Choi, Y.~Lee, B.-Y. Kim, and S.~Watanabe,
  ``{TF-GridNet}: Integrating full-and sub-band modeling for speech
  separation,'' \emph{IEEE/ACM Trans. Audio, Speech, Lang. Process.}, vol.~31,
  pp. 3221--3236, 2023.

\bibitem{stoter19}
F.-R. St{\"o}ter, S.~Uhlich, A.~Liutkus, and Y.~Mitsufuji, ``Open-{U}nmix - a
  reference implementation for music source separation,'' \emph{Journal of Open
  Source Software}, 2019.

\bibitem{sawata2021all}
R.~Sawata, S.~Uhlich, S.~Takahashi, and Y.~Mitsufuji, ``All for one and one for
  all: Improving music separation by bridging networks,'' in \emph{Proc.
  ICASSP}, Jun. 2021, pp. 51--55.

\bibitem{bsrnn}
Y.~Luo and J.~Yu, ``Music source separation with band-split {RNN},''
  \emph{IEEE/ACM Trans. Audio, Speech, Lang. Process.}, vol.~31, pp.
  1893--1901, 2023.

\bibitem{universal_sound_separation}
I.~Kavalerov, S.~Wisdom, H.~Erdogan, B.~Patton, K.~Wilson, J.~Le~Roux, and
  J.~R. Hershey, ``Universal sound separation,'' in \emph{Proc. WASPAA}, 2019,
  pp. 175--179.

\bibitem{tzinis2020improving}
E.~Tzinis, S.~Wisdom, J.~R. Hershey, A.~Jansen, and D.~P. Ellis, ``Improving
  universal sound separation using sound classification,'' in \emph{Proc.
  ICASSP}, 2020, pp. 96--100.

\bibitem{zhang2021multi}
L.~Zhang, C.~Li, F.~Deng, and X.~Wang, ``Multi-task audio source separation,''
  in \emph{Proc. ASRU}, 2021, pp. 671--678.

\bibitem{petermann2022cocktail}
D.~Petermann, G.~Wichern, Z.-Q. Wang, and J.~Le~Roux, ``The cocktail fork
  problem: Three-stem audio separation for real-world soundtracks,'' in
  \emph{Proc. ICASSP}, 2022, pp. 526--530.

\bibitem{Uhlich2024CDX}
S.~Uhlich, G.~Fabbro, M.~Hirano, S.~Takahashi, G.~Wichern, J.~Le~Roux,
  D.~Chakraborty, S.~Mohanty, K.~Li, Y.~Luo \emph{et~al.}, ``The sound demixing
  challenge 2023 – cinematic demixing track,'' \emph{Transactions of the
  International Society for Music Information Retrieval}, Apr. 2024.

\bibitem{pons2024gass}
J.~Pons, X.~Liu, S.~Pascual, and J.~Serr{\`a}, ``{GASS}: Generalizing audio
  source separation with large-scale data,'' in \emph{Proc. ICASSP}, 2024, pp.
  546--550.

\bibitem{vzmolikova2019speakerbeam}
K.~{\v{Z}}mol{\'\i}kov{\'a}, M.~Delcroix, K.~Kinoshita, T.~Ochiai, T.~Nakatani,
  L.~Burget, and J.~{\v{C}}ernock{\`y}, ``Speakerbeam: Speaker aware neural
  network for target speaker extraction in speech mixtures,'' \emph{IEEE J.
  Sel. Top. Signal Process.}, vol.~13, no.~4, pp. 800--814, 2019.

\bibitem{wang2022few}
Y.~Wang, D.~Stoller, R.~M. Bittner, and J.~P. Bello, ``Few-shot musical source
  separation,'' in \emph{Proc. ICASSP}, 2022, pp. 121--125.

\bibitem{chen2022zero}
K.~Chen, X.~Du, B.~Zhu, Z.~Ma, T.~Berg-Kirkpatrick, and S.~Dubnov, ``Zero-shot
  audio source separation through query-based learning from weakly-labeled
  data,'' in \emph{Proceedings of the AAAI Conference on Artificial
  Intelligence}, vol.~36, no.~4, 2022, pp. 4441--4449.

\bibitem{wang19h_interspeech}
Q.~Wang, H.~Muckenhirn, K.~Wilson, P.~Sridhar, Z.~Wu, J.~R. Hershey, R.~A.
  Saurous, R.~J. Weiss, Y.~Jia, and I.~L. Moreno, ``Voicefilter: Targeted voice
  separation by speaker-conditioned spectrogram masking,'' in \emph{Proc.
  Interspeech}, 2019, pp. 2728--2732.

\bibitem{seetharaman2019class}
P.~Seetharaman, G.~Wichern, S.~Venkataramani, and J.~Le~Roux,
  ``Class-conditional embeddings for music source separation,'' in \emph{Proc.
  ICASSP}, 2019, pp. 301--305.

\bibitem{ochiai20_interspeech}
T.~Ochiai, M.~Delcroix, Y.~Koizumi, H.~Ito, K.~Kinoshita, and S.~Araki,
  ``Listen to what you want: Neural network-based universal sound selector,''
  in \emph{Proc. Interspeech}, 2020, pp. 1441--1445.

\bibitem{tzinis2022heterogeneous}
E.~Tzinis, G.~Wichern, A.~Subramanian, P.~Smaragdis, and J.~Le~Roux,
  ``Heterogeneous target speech separation,'' in \emph{Proc. Interspeech},
  2022, pp. 1796--1800.

\bibitem{delcroix2022soundbeam}
M.~Delcroix, J.~B. V{\'a}zquez, T.~Ochiai, K.~Kinoshita, Y.~Ohishi, and
  S.~Araki, ``{SoundBeam}: Target sound extraction conditioned on sound-class
  labels and enrollment clues for increased performance and continuous
  learning,'' \emph{IEEE/ACM Trans. Audio, Speech, Lang. Process.}, vol.~31,
  pp. 121--136, 2022.

\bibitem{kilgour22_interspeech}
K.~Kilgour, B.~Gfeller, Q.~Huang, A.~Jansen, S.~Wisdom, and M.~Tagliasacchi,
  ``Text-driven separation of arbitrary sounds,'' in \emph{Proc. Interspeech},
  2022, pp. 5403--5407.

\bibitem{liu22w_interspeech}
X.~Liu, H.~Liu, Q.~Kong, X.~Mei, J.~Zhao, Q.~Huang, M.~D. Plumbley, and
  W.~Wang, ``Separate what you describe: Language-queried audio source
  separation,'' in \emph{Proc. Interspeech}, 2022, pp. 1801--1805.

\bibitem{saijo2023single}
K.~Saijo, W.~Zhang, Z.-Q. Wang, S.~Watanabe, T.~Kobayashi, and T.~Ogawa, ``A
  single speech enhancement model unifying dereverberation, denoising, speaker
  counting, separation, and extraction,'' in \emph{Proc. ASRU}, 2023, pp. 1--6.

\bibitem{manilow2020hierarchical}
E.~Manilow, G.~Wichern, and J.~Le~Roux, ``Hierarchical musical instrument
  separation.'' in \emph{ISMIR}, 2020, pp. 376--383.

\bibitem{petermann2023hyperbolic}
D.~Petermann, G.~Wichern, A.~Subramanian, and J.~Le~Roux, ``Hyperbolic audio
  source separation,'' in \emph{Proc. ICASSP}, 2023, pp. 1--5.

\bibitem{zhang2024urgent}
W.~Zhang, R.~Scheibler, K.~Saijo, S.~Cornell, C.~Li, Z.~Ni, J.~Pirklbauer,
  M.~Sach, S.~Watanabe, T.~Fingscheidt \emph{et~al.}, ``Urgent challenge:
  Universality, robustness, and generalizability for speech enhancement,'' in
  \emph{Proc. Interspeech}, 2024, pp. 4868--4872.

\bibitem{fuss}
S.~Wisdom, H.~Erdogan, D.~P. Ellis, R.~Serizel, N.~Turpault, E.~Fonseca,
  J.~Salamon, P.~Seetharaman, and J.~R. Hershey, ``What’s all the fuss about
  free universal sound separation data?'' in \emph{Proc. ICASSP}, 2021, pp.
  186--190.

\bibitem{veaux2013voice}
C.~Veaux, J.~Yamagishi, and S.~King, ``The {V}oice {B}ank corpus: Design,
  collection and data analysis of a large regional accent speech database,'' in
  \emph{Proc. O-COCOSDA/CASLRE}, 2013, pp. 1--4.

\bibitem{wsj0}
J.~S. Garofolo \emph{et~al.}, \emph{{CSR-I} ({WSJ0}) Complete {LDC93S6A}},
  Linguistic Data Consortium, Philadelphia, 1993, web Download.

\bibitem{fonseca2021fsd50k}
E.~Fonseca, X.~Favory, J.~Pons, F.~Font, and X.~Serra, ``Fsd50k: an open
  dataset of human-labeled sound events,'' \emph{IEEE/ACM Trans. Audio, Speech,
  Lang. Process.}, vol.~30, pp. 829--852, 2021.

\bibitem{wichern19_interspeech}
G.~Wichern, J.~Antognini, M.~Flynn, L.~R. Zhu, E.~McQuinn, D.~Crow, E.~Manilow,
  and J.~Le~Roux, ``{WHAM}!: Extending speech separation to noisy
  environments,'' in \emph{Proc. Interspeech}, 2019, pp. 1368--1372.

\bibitem{thiemann2013diverse}
J.~Thiemann, N.~Ito, and E.~Vincent, ``The diverse environments multi-channel
  acoustic noise database ({DEMAND}): A database of multichannel environmental
  noise recordings,'' in \emph{Proc. Mtgs. Acoust.}, vol.~19, no.~1, 2013.

\bibitem{MUSDB18HQ}
\BIBentryALTinterwordspacing
Z.~Rafii, A.~Liutkus, F.-R. St{\"o}ter, S.~I. Mimilakis, and R.~Bittner,
  ``{MUSDB18-HQ} - an uncompressed version of {MUSDB18},'' Dec. 2019. [Online].
  Available: \url{https://doi.org/10.5281/zenodo.3338373}
\BIBentrySTDinterwordspacing

\bibitem{pereira2023moisesdb}
I.~Pereira, F.~Ara{\'u}jo, F.~Korzeniowski, and R.~Vogl, ``Moises{DB}: A
  dataset for source separation beyond 4-stems,'' \emph{arXiv preprint
  arXiv:2307.15913}, 2023.

\bibitem{defferrard2016fma}
M.~Defferrard, K.~Benzi, P.~Vandergheynst, and X.~Bresson, ``{FMA}: A dataset
  for music analysis,'' \emph{arXiv preprint arXiv:1612.01840}, 2016.

\bibitem{tflocoformer}
K.~Saijo, G.~Wichern, F.~G. Germain, Z.~Pan, and J.~Le~Roux, ``{TF-Locoformer}:
  Transformer with local modeling by convolution for speech separation and
  enhancement,'' \emph{arXiv preprint arXiv:2408.03440}, 2024.

\bibitem{bsroformer}
W.-T. Lu, J.-C. Wang, Q.~Kong, and Y.-N. Hung, ``Music source separation with
  band-split rope transformer,'' in \emph{Proc. ICASSP}, 2024, pp. 481--485.

\bibitem{adamw}
I.~Loshchilov and F.~Hutter, ``Decoupled weight decay regularization,'' in
  \emph{Proc. ICLR}, 2018.

\bibitem{zhang2024beyond}
W.~Zhang, K.~Saijo, J.~weon Jung, C.~Li, S.~Watanabe, and Y.~Qian, ``Beyond
  performance plateaus: A comprehensive study on scalability in speech
  enhancement,'' in \emph{Proc. Interspeech}, 2024, pp. 1740--1744.

\end{thebibliography}

\end{document}